\documentclass[%
 reprint,
superscriptaddress,
nofootinbib,
 amsmath,amssymb,
 aps,
prl,
]{revtex4-2}

\usepackage{graphicx}
\usepackage{dcolumn}
\usepackage{bm}
\usepackage{color}

\begin{document}

\title{Strain control of the metal-insulator transition in epitaxial SrCrO$_{3}$ thin films}

\author{Giulia Bertino}
 \affiliation{Department of Physics and Astronomy, Stony Brook University, Stony Brook, New York 11794-3800, USA}
 \author{Hsiang-Chun Hsing}%
 \affiliation{Department of Physics and Astronomy, Stony Brook University, Stony Brook, New York 11794-3800, USA}
 \author{Anna Gura}%
 \affiliation{Department of Physics and Astronomy, Stony Brook University, Stony Brook, New York 11794-3800, USA}
 \author{Xinzhong Chen}%
 \affiliation{Department of Physics and Astronomy, Stony Brook University, Stony Brook, New York 11794-3800, USA}
\author{Theodore Sauyet}%
 \affiliation{Department of Physics and Astronomy, Stony Brook University, Stony Brook, New York 11794-3800, USA}
\author{Mengkun Liu}%
 \affiliation{Department of Physics and Astronomy, Stony Brook University, Stony Brook, New York 11794-3800, USA}
\author{Chang-Yong Nam}%
 \affiliation{Center for Functional Nanomaterials, Brookhaven National Laboratory, Upton, New York 11973, USA}
\author{Matthew Dawber}
 \affiliation{Department of Physics and Astronomy, Stony Brook University, Stony Brook, New York 11794-3800, USA}

\begin{abstract}
In the perovskite oxide SrCrO$_{3}$ the interplay between crystal structure, strain and orbital ordering enables a transition from a metallic to an insulating electronic structure under certain conditions. We identified a narrow window of oxygen partial pressure in which highly strained SrCrO$_{3}$ thin films can be grown using radio-frequency (RF) off-axis magnetron sputtering on three different substrates, (LaAlO$_{3}$)$_{0.3}$-(Sr$_{2}$TaAlO$_{6}$)$_{0.7}$ (LSAT), SrTiO$_{3}$ (STO) and DyScO$_{3}$ (DSO). X-ray diffraction and atomic force microscopy confirmed the quality of the films and a metal-insulator transition driven by the substrate induced strain was demonstrated.

\end{abstract}

\maketitle

Transition metal oxides with perovskite structure ABO$_{3}$ attract great interest thanks to the novel properties that these materials can achieve due to the interplay among strain, lattice deformation, spin and electronic structure. Key to their properties is the partially occupied $d$ shell, whose levels are generally divided in two main groups, $t_{2g}$ and $e_{g}$, due to crystal field splitting. The first splits into $d_{xy}, d_{xz}$ and $d_{yz}$ levels, where $d_{xz}$ and $d_{yz}$ are degenerate, while the second splits into $d_{3z^{2}-r^2}$ and $d_{x^2-y^2}$ levels \cite{imada1998metal}. When an external source like strain or defects deforms the crystal structure, the distortion is always accompanied by a rearrangement of the electronic levels, which, in turn, causes the opening of a gap inside the $d$ shell: this entire process is commonly referred to as orbital ordering and it characterizes transition metals defined as Mott insulators \cite{mott1937discussion, mott1949basis, mott1956transition, mott1961transition, mott1968conduction, mott1974metal, zaanen1985band, sawatzky1984magnitude, hufner1985electronic}. Orbital ordering within the partially occupied $d$ levels in these materials can be exploited to engineer a metal-insulator transition (MIT) that depends on strain. Epitaxial SrCrO$_{3}$ (SCO) thin films are a strong candidate to achieve such a strain-controlled MIT, though an additional complication is the extremely strong dependence of the properties of a film on oxygen stochiometry, which so far has prevented complete and systematic experimental studies aimed at exploring this effect \citep{zhang2014reversible,zhang2015electronic}.

\begin{figure}[ht]
\includegraphics[width=8.5cm]{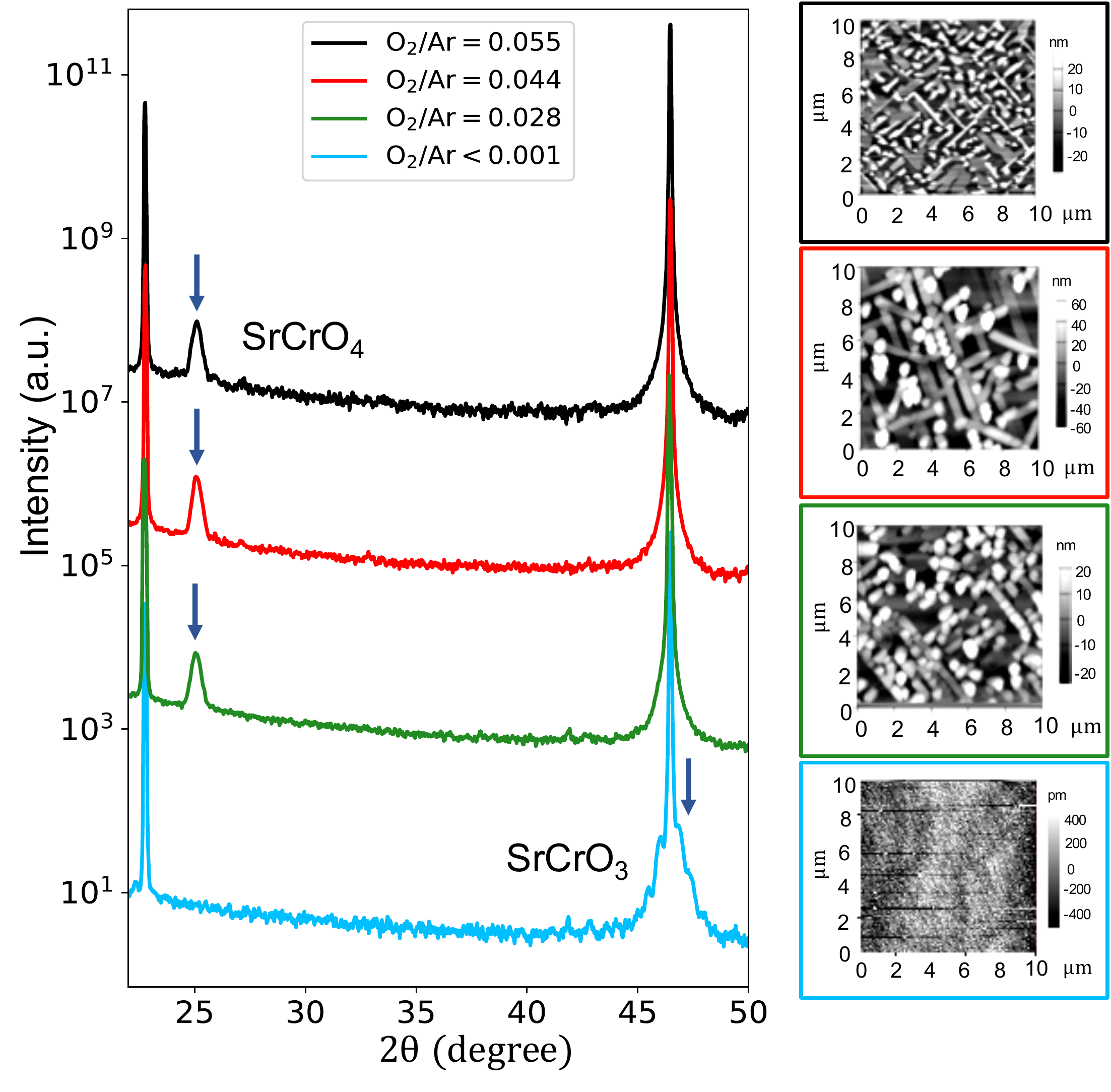}
\caption{\textbf{Growth conditions to obtain the correct phase of SCO thin films on STO.} On the left column, the XRD spectra show the characteristic peak of the oxygen-rich phase SrCrO$_{4}$, which is obtained for O$_{2}$/Ar$\geq$0.028. Only when O$_{2}$/Ar$<$0.001 the correct SrCrO$_{3}$ phase is obtained. On the right column, the AFM images of the corresponding sample surfaces are shown: SrCrO$_{4}$ surfaces are dominated by large crystals and do not show a uniform film layer. The crystals decrease in size with the decrease of O$_{2}$/Ar ratio, until a uniform film layer is reached when the correct phase is achieved.} 
\label{RightPhase}
\vspace{-4mm}
\end{figure}

Bulk SCO was first reported to be a paramagnetic metal with cubic structure in 1967 \citep{chamberland1967preparation}. However, more recent works on polycrystalline bulk samples report a pressure-driven MIT due to Cr$-$O bond length instability \citep{zhou2006anomalous} and a phase coexistence of cubic and tetragonal phase in a low temperature range ($35-70$ K), with both phases being metallic \citep{ortega2007microstrain}. 

From the perspective of first principles calculations, Lee and Pickett found that in bulk SCO the tetragonal phase undergoes an orbital ordering $t_{2g}\rightarrow d_{xy}^{1}(d_{xz}, d_{yz})^{1}$ due to a lattice distortion when the Coulomb $U$ term exceeds 4 eV \cite{lee2009orbital}, but the degeneracy of $d_{xz}$ and $ d_{yz}$ was maintained and with it the metallicity of the material. Qian et al. also arrived at similar conclusions in their study, though they did find that at unrealistically large values of $U$ the $d_{xz}$ and $ d_{yz}$ orbitals could become lower in energy than $d_{xy}$, which led to an insulating state \cite{qian2011electronic}.

\begin{figure}[ht]
\includegraphics[width=0.337\textwidth]{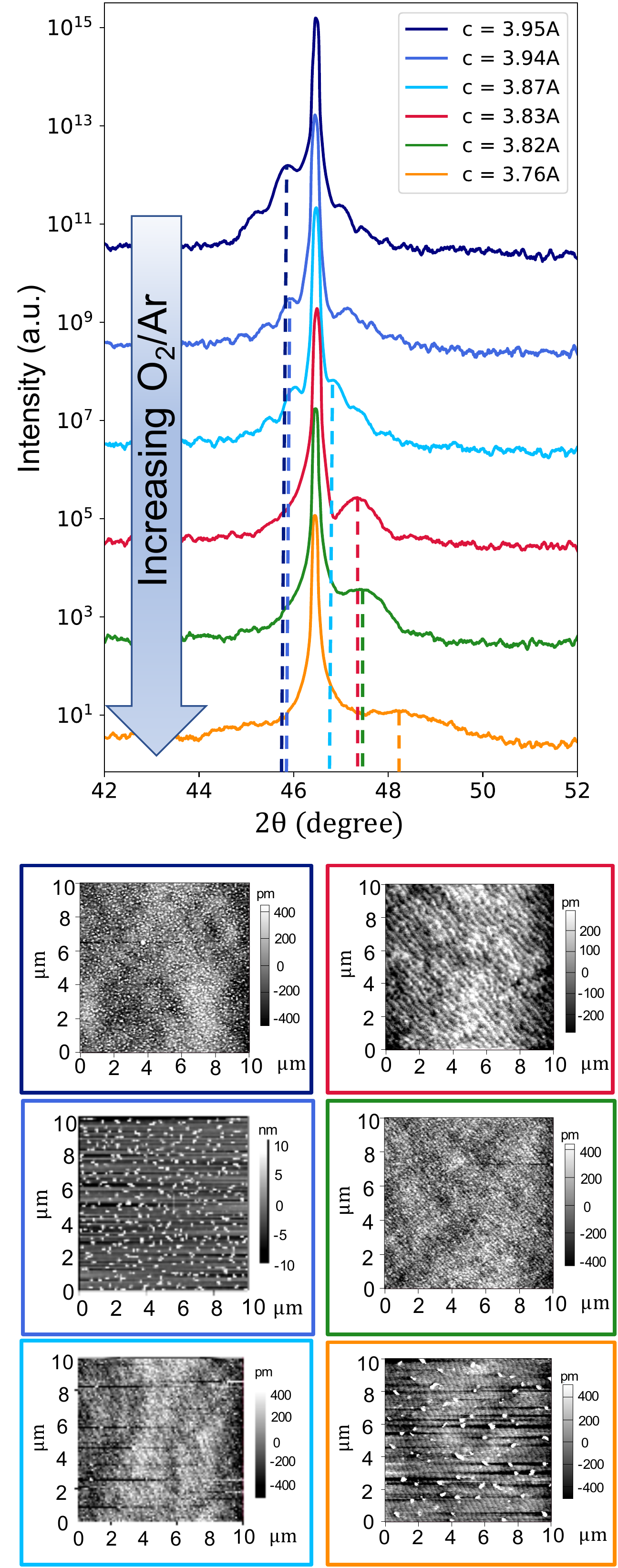}
\caption{\textbf{The effect of the variation of O$_{2}$/Ar on the lattice parameter and the surface quality.} Top: XRD spectra of SCO films grown with different O$_{2}$/Ar ratio: when O$_{2}$/Ar$<0.0005$ the film is oxygen deficient, the film peak is at the left of the STO peak and the lattice parameter is larger than the STO substrate. This effect is due to the chemical expansivity phenomenon \cite{adler2001chemical}. When $0.0005<$O$_{2}$/Ar$<0.001$, the SCO peak moves to the right of the STO peak and the lattice parameter slightly changes as the oxygen partial pressure increases. Bottom: AFM surface images of each film: it is clear that the surface morphology strongly improves when $0.0005<$O$_{2}$/Ar$<0.001$.}
\label{RightO2}
\vspace{-4mm}
\end{figure}

To produce a change of ordering from $t_{2g}\rightarrow d_{xy}^{1}(d_{xz}, d_{yz})^{1}$ (metallic) to $t_{2g}	\rightarrow (d_{xz}, d_{yz})^{2}d_{xy}^{0}$ (insulator), a combination of requirements have been established from first principles. Gupta et al. \citep{gupta2013orbital} showed with a first principles calculation that a SCO thin film of two unit cells on SrTiO$_{3}$ (STO) (tensile strain 2\%) substrate is in an \textit{insulating} state with a finite electrical polarization. This state is enabled by the missing apical oxygen of the Cr-centered oxygen octahedron at the surface. Zhou and Rabe \cite{zhou2015coupled} showed that modification of the $d$ band states of SCO by combining it with STO in a 1/1 superlattice would also enable $t_{2g}	\rightarrow (d_{xz}, d_{yz})^{2}d_{xy}^{0}$ ordering when tensile strain imposed on the superlattice exceeded 2.2$\%$ which would stabilize an in-plane polar distortion. These results motivated us to pursue the growth of epitaxial thin films of SCO at very high tensile strains to attempt to realize the insulating phase of SCO.  

Experimentally, Zhang et al. and Ong et al. \cite{zhang2014reversible, zhang2015electronic, ong2017low} showed that $50$ nm epitaxial  thin films of SrCrO$_{3-\delta}$ on LaAlO$_{3}$ (LAO) (compressive strain $-0.7\%$) and (LaAlO$_{3}$)$_{0.3}$-(Sr$_{2}$TaAlO$_{6}$)$_{0.7}$ (LSAT) (tensile strain $1.2\%$) show a reversible transformation with oxygen content between a rhombohedral \textit{semiconductor} phase (R-SCO) and a cubic \textit{metallic} phase (P-SCO), which is accompanied by a rearrangement of oxygen vacancies in layers of [111]-oriented planes and a change of the valence of Cr atoms from Cr$^{4+}$ to Cr$^{3+}$(the last valence state being prevalent in P-SCO). The segregation of oxygen vacancies into superstructures of [111]-oriented layers and the change of the Cr atoms valence was also previously reported  for polycrystalline bulk SCO  \cite{arevalo2012hard} and it has been demonstrated that oxygen diffusion through the films happens fastest through these superstructures \cite{zhang2014reversible}. However, the experimental demonstration of strain-induced MIT driven by orbital ordering has not been reported so far because the largest value of epitaxial strain that has been achieved to date ($1.2\%$ tensile strain on LSAT \cite{zhang2014reversible, zhang2015electronic}) is not large enough to drive the system toward an insulating phase.

In this paper, we report the successful growth of SCO thin films on $(001)$ LSAT, $(001)$ STO and $(110)$ DyScO$_{3}$ (DSO), which impose tensile strains of 1.2\%, 2\% and 3.2\% on the films, respectively, calculated with respect to the bulk lattice parameter of SCO. Successful growth on STO and DSO are significant advances, as the maximum strain previously reached experimentally in literature is only 1.2\% on LSAT from Zhang et al. \cite{zhang2014reversible,zhang2015electronic}, well below the limits we were able to reach in the present study. We characterize the surfaces of films through atomic force microscopy (AFM) topography images and perform a thorough crystal and electric characterization through X-Ray Diffraction (XRD), in-situ synchrotron studies, reciprocal space maps and the van der Pauw method. We induce a transition in the electrical properties by changing the epitaxial strain imposed on SCO thin films by the different substrates and confirm a transition from a metallic to an insulating state at increasingly high tensile strains.

Thin films were grown using $90^\circ$ off-axis radio-frequency (RF) magnetron sputtering from ceramic SCO targets on single crystal substrates (supplied by CrysTec). The deposition atmosphere was mostly argon with small amounts of oxygen. Characterizations of the crystal quality through 2$\theta/\omega$ XRD scans around the (002) peak of all substrates were performed with a Bruker D8 Discover diffractometer,  which  uses  copper K$\alpha1$ radiation ($\lambda= 1.5402$ \AA). To ensure that all our samples are epitaxially strained on the substrates we performed reciprocal space mapping around the (101) substrate peaks at the 4-1D beamline of National Synchrotron Light Source II (NSLS-II) at Brookhaven National Laboratory ($\lambda = 1.0783$ \AA). We also evaluate the surface quality of the films through AFM images.

Little information is present in literature regarding the growth of SCO through sputtering. It is known from the growth studies using molecular beam epitaxy that a successful growth of SCO typically requires high temperature and pressure \cite{zhou2006anomalous, ortega2007microstrain, arevalo2012hard, zhang2014reversible, zhang2015electronic}. We started experimenting by growing 34 nm-thick films of SCO on STO via sputtering and tuned the oxygen partial pressure to find the optimal growth conditions. Specifically, we started by using an overall pressure inside the growth chamber of 24 mbar, a growth temperature of $700^\circ$C and by varying the Ar and O$_{2}$ flow rate ratio to achieve the correct stoichiometry in the films. This last step is crucial to achieve an optimal growth for SCO thin films.

\begin{figure}[t]
\includegraphics[width=0.5\textwidth]{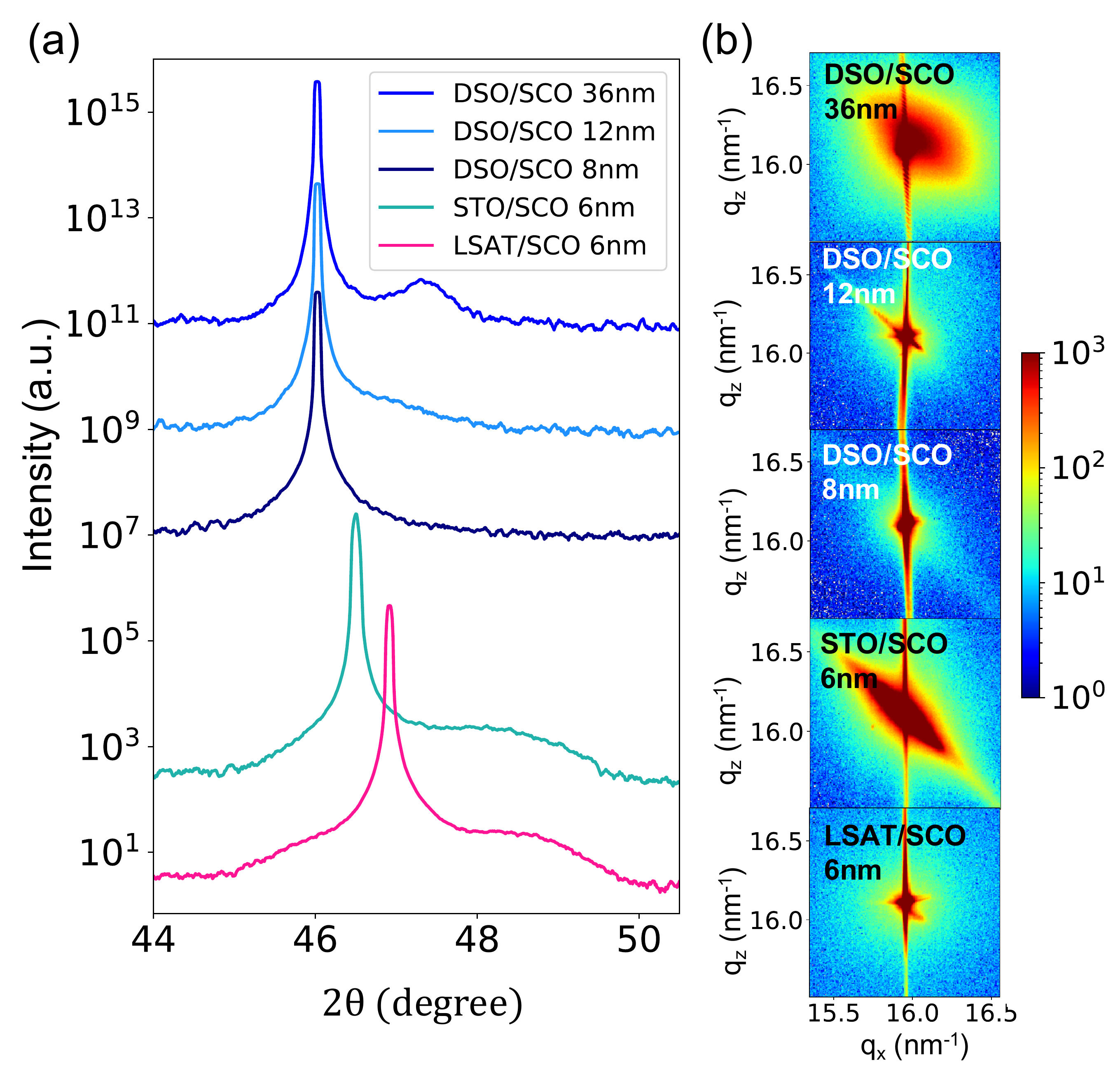}
\caption{\textbf{SCO films on different substrates.} (a) XRD spectra of thin SCO films on the different substrates: LSAT, STO and DSO. (b) Reciprocal space maps around the (101) substrate peak for all the SCO films. The 36 nm-thick film on DSO shows a partial relaxation, while the others appear fully epitaxially strained to in-plane substrate lattice parameter.}
\label{DiffSubstrates}
\vspace{-4mm}
\end{figure}

Indeed, we find SCO is extremely sensitive to the amount of oxygen present during the growth, to the point that a minimal fluctuation in the O$_{2}$ flow leads to dramatic changes both in the stoichiometry and the surface morphology of the samples. At a ratio of O$_{2}$/Ar$= 0.055$, the XRD spectra reveal a peak at $25.10^\circ$ typical of the oxygen-rich phase SrCrO$_{4}$, as shown in the black spectrum in Fig. \ref{RightPhase}. Lowering to oxygen ratio to O$_{2}$/Ar$= 0.044$ and O$_{2}$/Ar$= 0.028$ still produces the oxygen-rich phase of the films (red and green spectra in Fig. \ref{RightPhase}). Finally, when O$_{2}$/Ar is reduced below 0.001, we were able to obtain the correct SrCrO$_{3}$ phase with the characteristic peak at $46.83^\circ$, as shown by the light blue spectrum in Fig. \ref{RightPhase}.
At the same time, the surface morphology of the films dramatically changes with O$_{2}$/Ar: as shown in the right column of Fig. \ref{RightPhase}, the corresponding AFM images of the oxygen-rich SrCrO$_{4}$ samples show a surface characterized by large crystals several nanometers high and a non-uniform rough structure. However, these crystals decrease in size as O$_{2}$/Ar ratio decreases and once the correct stoichiometry is reached, the surface appears smooth, with a clear step and terrace structure that resembles that of the substrate.

Within the correct phase, we find that the amount of oxygen needed to achieve a good quality SrCrO$_{3}$ must be controlled in the very small window of $0.0005<$O$_{2}$/Ar$<0.001$. Indeed, if the O$_{2}$/Ar ratio is reduced to slightly less than 0.0005, the peak of the film moves to the left of the STO substrate peak, as shown in Fig. \ref{RightO2} by the top two spectra. These two films, although still in the correct phase, are oxygen deficient and the surface quality quickly degrades with the appearance of small islands as seen in the AFM images on the bottom of Fig. \ref{RightO2}.

Small excesses in oxygen can also lead to problems, but these are harder to identify as they do not lead to the same changes in surface topography.  Fig. \ref{RightO2} shows that films grown with slightly higher O$_{2}$/Ar ratio (the yellow and green curves) have smaller out-of-plane lattice parameters, but that the surface only begins to degrade in the case of the film with \textit{c} axis lattice parameter of 3.76 \AA, and this degradation is much less severe than in the case of oxygen deficient samples.

The film quality could be further confirmed by growing SCO/STO superlattices (SL) on STO. We grew the STO layers using the same growth conditions as SCO. While small changes in the O$_{2}$/Ar ratio  allow the growth of SCO films with different lattice parameters, not all of them allow the growth of superlattices. The conditions used to obtain the final two films shown in Fig. \ref{RightO2}, with lattice parameters $c= 3.76$ \AA~ and $c= 3.82$ \AA~ do not lead to successful superlattices. However, when the ratio O$_{2}$/Ar$<0.001$ was used to produce the film with $c= 3.83$ \AA, we do obtain successful superlattices (Fig. 1 in Supplemental Material), which have not previously been reported in the literature.

\begin{figure}[t]
\includegraphics[width=0.4\textwidth]{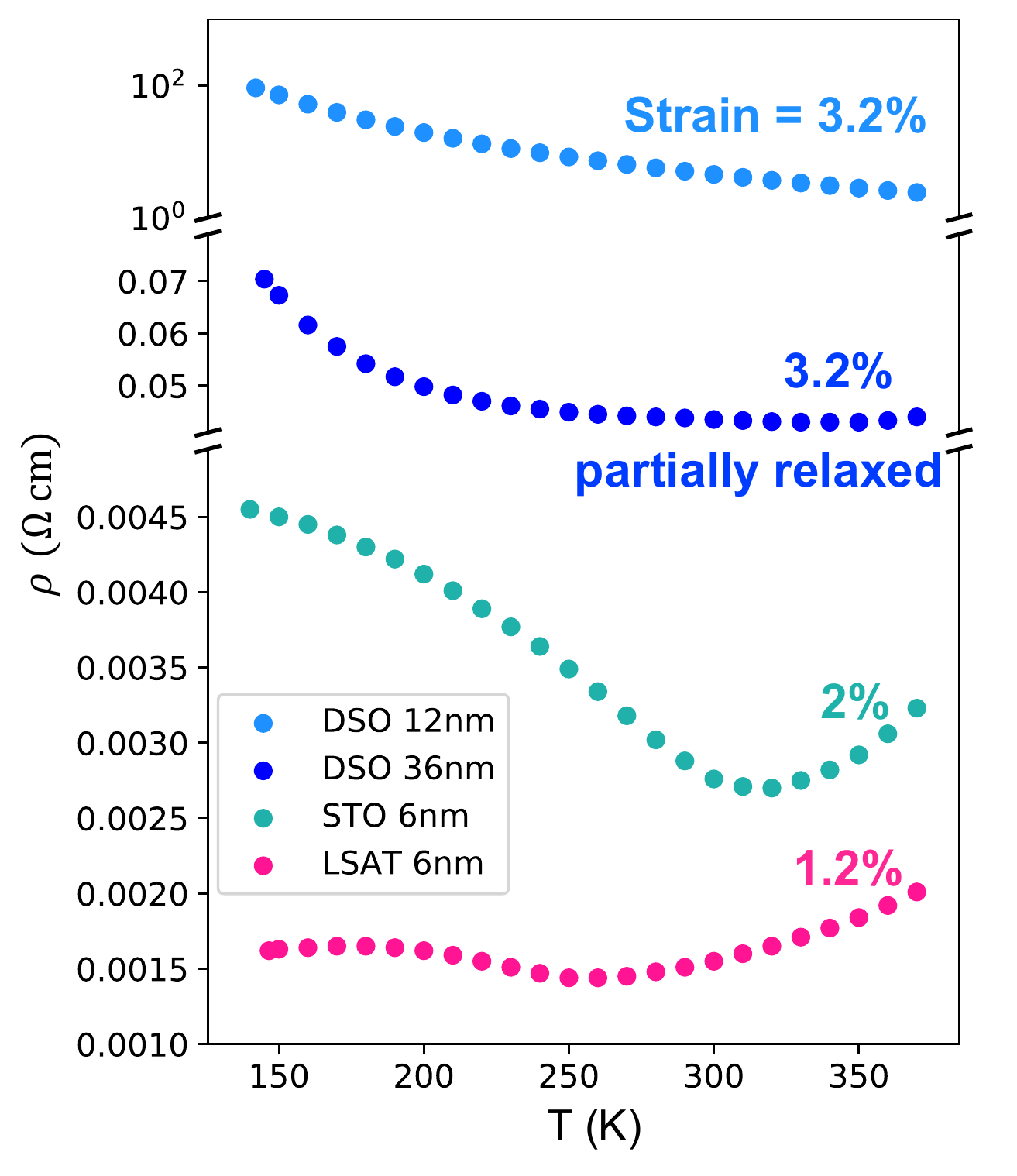}
\caption{\textbf{Resistivity characteristics as a function of temperature for SCO thin films.} Resistivity versus temperature characteristics measured with the van der Pauw method at the Center for Functional Nanomaterials (CFN) for the SCO films on the different substrates. The 8 nm film on DSO is too resistive to be measured with this technique, hence no data are reported for it. While the films on LSAT and STO show a metallic behavior since $\partial\rho/\partial T>0$ starting at 250 K and 310 K, respectively, the films on DSO show a transition to an insulating behavior, since $\partial\rho/\partial T<0$. This proves that a metal-insulator transition is achieved by tuning the strain on SCO thin films and that larger strains drive the films from a metallic to an insulating characteristic.}
\label{Resistivity} 
\vspace{-4mm}
\end{figure}

Having found the optimal O$_{2}$/Ar ratio, we synthesized samples on multiple substrates at the 4-ID beamline at NSLS-II, which allows in-situ XRD to be performed during growth. Here, we used an overall pressure of 33 mbar, a growth temperature of 700$^{\circ}$C and we applied a sputtering power of 40 W to each target. By successfully and consistently growing SCO/STO superlattices on STO we found an optimal ratio of O$_{2}$/Ar$=0.00057$, which is consistent with the window previously identified of $0.0005<$O$_{2}$/Ar$<0.001$. These parameters produced SCO films on STO with a $c$ lattice parameter of 3.78 \AA.

Thin films of SCO were  grown, all under the aforementioned conditions, on the three different substrates, LSAT, STO and DSO, which apply strains of 1.2\%, 2\% and 3.2\%, respectively, on the SCO films atop. The results are shown in Fig. \ref{DiffSubstrates} (a). The lattice parameters of the 6 nm-thick films grown on LSAT and STO are $c=3.75$ \AA~ and $c=3.78$ \AA, respectively. 

On DSO we grew films of three different thicknesses, 8 nm, 12 nm and 36 nm. The first two have a out-of-plane lattice parameter equal to the substrate, $c=3.94$ \AA. The 36 nm film, on the other hand, has a lattice parameter $c=3.84$ \AA. The successful growth of these films were confirmed by the oscillations of anti-Bragg peak (0\,0\,1/2) intensity observed the during the growth (Fig. 2 in Supplemental Materials). Meanwhile, the (101) diffraction peak becomes most prominent when the film thickness reaches 36 nm. This is in fact a remarkable result, as the substrates previously used in literature are only LAO (compressive strain of $-0.7$\%) and LSAT \cite{zhang2014reversible, zhang2015electronic}, and this is, to the best of our knowledge, the first experimental result of the growth of SCO thin films under very high strain exceeding 2\% and over to 3\%, using STO and, more importantly, DSO substrates. Fig. \ref{DiffSubstrates} (b) shows the reciprocal space maps around the (101) substrate peak for all of the grown SCO films. It is clear that all of them, with the exception of the 36 nm-thick film on DSO, are full epitaxially constrained in-plane to the substrate, as no external features with respect to the central peak are present. The 36 nm film on DSO was grown for a longer time in order to obtain a relaxed film: the (101) reciprocal space map reveals, indeed, that the film is partially relaxed. The films grown on DSO have higher roughness than those on STO and LSAT (Fig. 3 in Supplemental Materials) most likely due to the higher mismatch strain. A major contributor to the roughness of the films are small particles on their surfaces which we believe to be SrO islands that form during the deposition and cooling process.

In order to test the conductive properties of our samples, with the ultimate goal of verifying the expected existance of a MIT, temperature-dependent sheet resistance ($\rho$) was measured in the temperature range of $140 - 370$ K using the Van der Pauw method, as shown in Fig. \ref{Resistivity}. Since the SCO films are extremely sensitive to high temperature annealings, which can irreversibly damage the films even in controlled oxygen atmosphere, we did not raise the temperature above 370 K. The results are shown in Fig. \ref{Resistivity}: on LSAT, where the strain is the smallest at 1.2\%, a metallic behavior, an increasing of the resistivity with the temperature $(\partial\rho/\partial T>0)$, starts at 250 K, while when we increase the strain to 2\% on STO this behavior appears at 310 K. However, when the strain is increased to 3.2\% on DSO the trend dramatically changes toward an insulating characteristic with decreasing resistivity with the temperature $(\partial\rho/\partial T<0)$, along with a significant increase in the magnitude of the resistivity. The 8 nm-thick film on DSO is too insulating to be characterized with this technique; consequently its resistivity would be higher than that of the 12 nm-thick film. Also, since the 36 nm-thick film on DSO is partially relaxed, it is expected to be more conductive than the strained 12 nm-thick film, which is indeed observed, although it does maintain the non-metallic behavior at low temperature (Fig. \ref{Resistivity}). The insulating phase observed under the increased strain in this study appears to be compatible with that predicted in the previous theoretical studies on strained thin films and superlattices \citep{gupta2013orbital, zhou2015coupled}; we expect that it is due to the reordering of the $d$ levels. In this study we did not explore the importance of  having a free surface or superlattice boundary conditions in enabling the transition, but as we have demonstrated the ability to fabricate superlattices we will explore the latter in a future study.

To summarize the importance of our work; theoretical studies had predicted a strain-induced transition for this material, but the challenges deriving from the required tensile strains at which this phenomenon was expected to take place had prevented its realization, as experiments prior to this work had only reached a tensile strain of 1.2\%. In contrast, we were able to push the strain limit beyond the theoretical threshold, to 3.2\%, by growing epitaxially strained SCO thin films on STO and DSO and demonstrated experimentally the existence of a strain induced MIT in SCO thin films. In doing so we have also identified the optimal oxygen atmosphere window for growth, which led us to a successful growth of SCO/STO superlattices for the first time. This will ultimately allow much finer tuning of the electronic structure, so that we can control the location of the metal-insulator transition of the material with respect to strain and temperature.

The first potential application that comes to mind for a strain induced MIT is a switching device. Films on the edge of such a transition could potentially be very effective at sensing local strains and converting them into electrical outputs. Further, Gupta et. al. \citep{gupta2013orbital} predicted using Monte Carlo simulations that SCO thin films in the insulating phase should exhibit a coupling between electrical and magnetic degrees of freedom, suggesting that the transition could be promoted by an applied magnetic field. We therefore see potential for films and superlattices containing strained SCO to be used as sensitive circuit elements that can rapidly respond to imposed strains and magnetic fields. 

\begin{acknowledgments}

This material is based upon work supported by the National Science Foundation under Grant Nos. DMR-1334867 and DMR-1506930.  This research used the beamline 4-ID of the National Synchrotron Light Source II (NSLS-II) and the resources at the Center for Functional Nanomaterials (CFN), both of which are U.S. Department of Energy (DOE) Office of Science User Facilities operated for the DOE Office of Science by Brookhaven National Laboratory under Contract No. DE-SC0012704.

\end{acknowledgments}

\end{document}


\title{SUPPLEMENTAL MATERIAL for: Strain control of the metal-insulator transition in epitaxial SrCrO$_{3}$ thin films}

\author{Giulia Bertino}
 \affiliation{Department of Physics and Astronomy, Stony Brook University, Stony Brook, New York 11794-3800, USA}
 \author{Hsiang-Chun Hsing}%
 \affiliation{Department of Physics and Astronomy, Stony Brook University, Stony Brook, New York 11794-3800, USA}
 \author{Anna Gura}%
 \affiliation{Department of Physics and Astronomy, Stony Brook University, Stony Brook, New York 11794-3800, USA}
 \author{Xinzhong Chen}%
 \affiliation{Department of Physics and Astronomy, Stony Brook University, Stony Brook, New York 11794-3800, USA}
\author{Theodore Sauyet}%
 \affiliation{Department of Physics and Astronomy, Stony Brook University, Stony Brook, New York 11794-3800, USA}
\author{Mengkun Liu}%
 \affiliation{Department of Physics and Astronomy, Stony Brook University, Stony Brook, New York 11794-3800, USA}
\author{Chang-Yong Nam}%
 \affiliation{Center for Functional Nanomaterials, Brookhaven National Laboratory, Upton, New York 11973, USA}
\author{Matthew Dawber}
 \affiliation{Department of Physics and Astronomy, Stony Brook University, Stony Brook, New York 11794-3800, USA}

\maketitle

\begin{figure}[h!]
\includegraphics[width=0.4\textwidth]{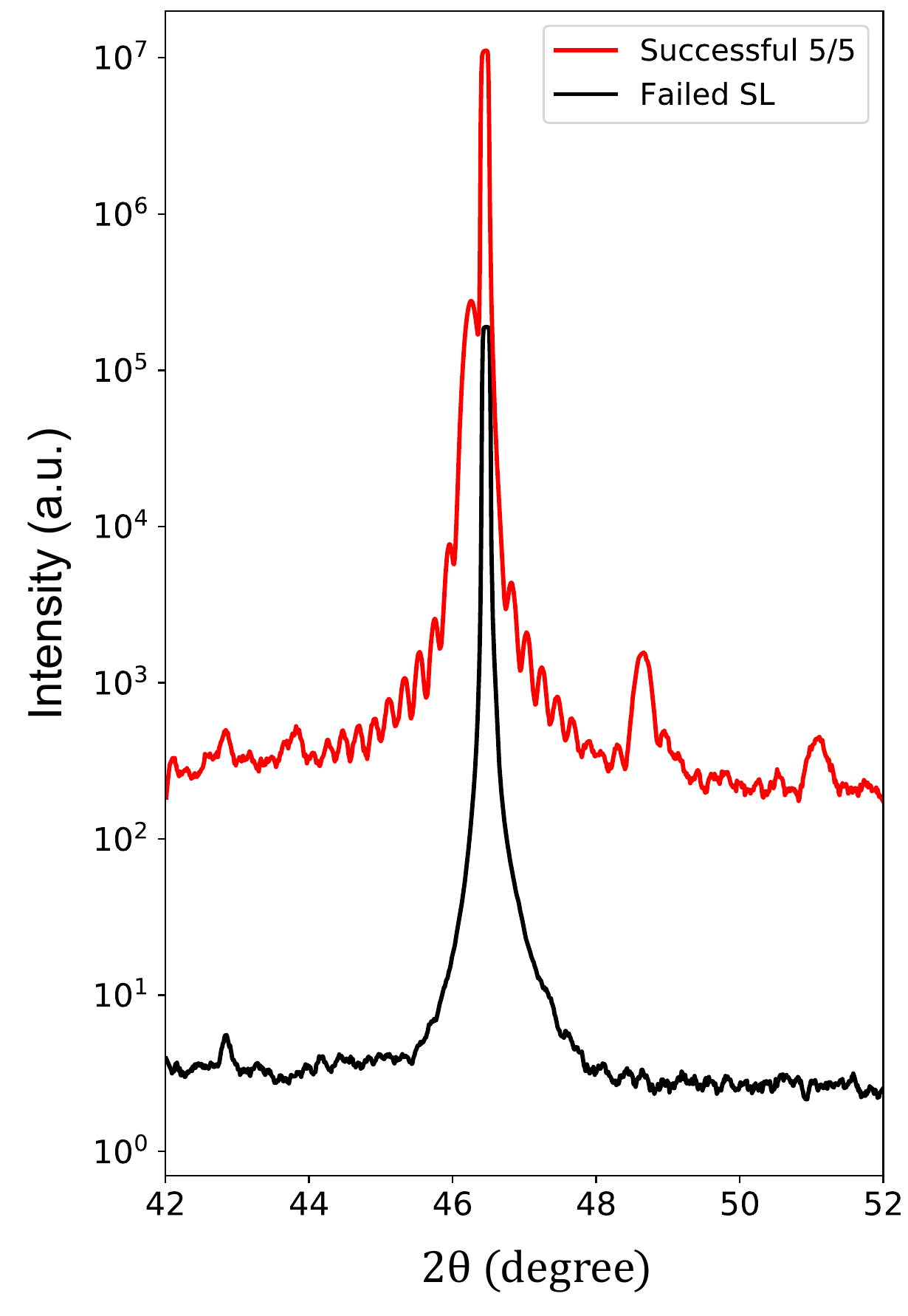}
\caption{\textbf{Successful and unsuccessful SCO/STO superlattices.} SCO/STO superlattices require a very specific ratio of O$_2$/Ar to enable the growth of the heterostructure. In this case, the black superlattice was grown in the same conditions of SCO film with out-of-plane lattice parameter $c=3.76$ \AA, while the red one corresponds to the film with $c=3.83$ \AA~ (see Fig. 2 in the main text). Both of these samples were grown with 0.0005$<$O$_2$/Ar$<$0.0006 with the successful superlattice having slightly higher oxygen than the unsuccessful one. }
\label{Superlattices}
\end{figure}

\begin{figure}[h!]
\includegraphics[width=0.6\textwidth]{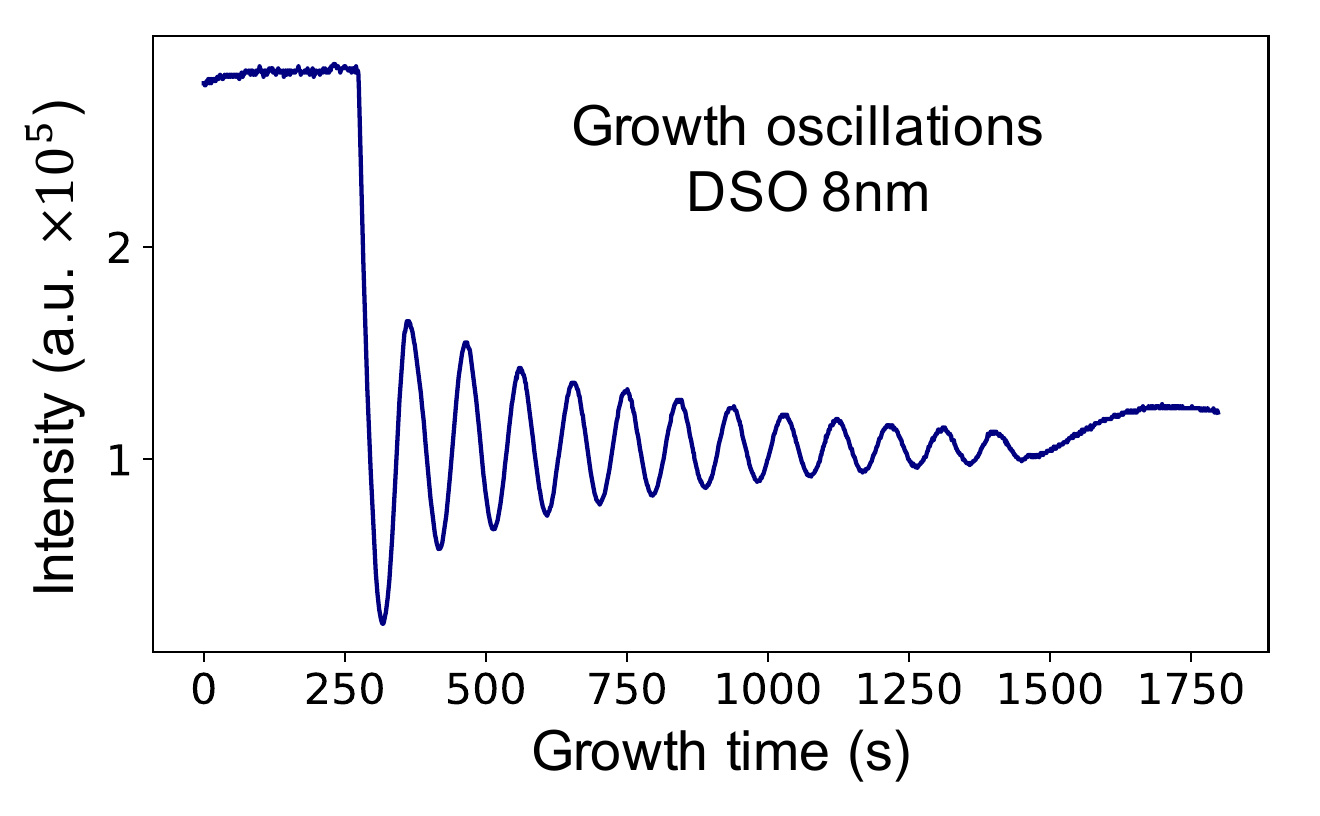}
\caption{\textbf{Intensity oscillations of the anti-Bragg peak (0\,0\,1/2) for the 8 nm SCO film on DSO.} The successful observation of the intensity oscillations during the entire growth process at the 4-1D beamline at NSLS II proves that the SCO films on DSO are effectively growing even in the presence of the extremely high tensile strain of 3.2\% on this substrate. The absence of the diffraction peak on the XRD spectra for the thin films on DSO (8 nm and 12 nm) is because the out of plane lattice parameter of the film is very close to the one of the substrate when it is epitaxially strained on DSO.}
\label{GrowthOscillations}
\end{figure}

\begin{figure}[h!]
\includegraphics[width=0.8\textwidth]{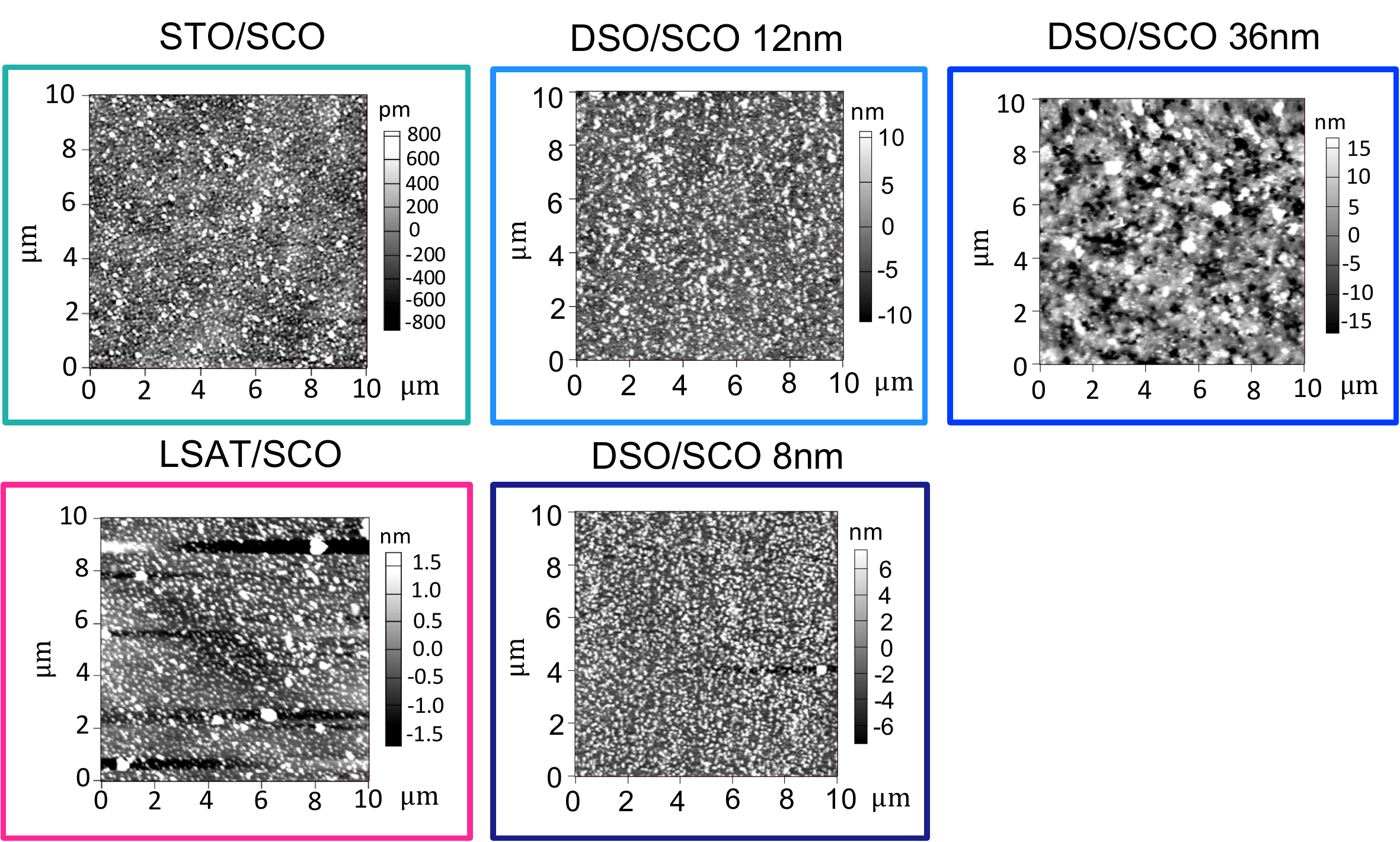}
\caption{\textbf{AFM images of the SCO films grown on LSAT, STO and DSO.} Films on DSO have higher roughness than those on STO and LSAT, most likely because of the larger mismatch strain. The root mean square (RMS) of the surfaces roughness is: 0.72 nm for LSAT/SCO, 0.36 nm for STO/SCO, 2.7 nm for DSO/SCO (8 nm thick), 3.3 nm for DSO/SCO (12 nm thick) and 5.7 nm for for DSO/SCO (36 nm thick). The increased roughness comes mainly from small particles that grow on the surface of the film that we believe to be SrO.}
\label{AFMfilms}
\end{figure}